\documentclass[aps,showpacs,pra,twocolumn]{revtex4}
\usepackage{graphicx}
\usepackage{amsmath}
\usepackage{wasysym}

\newcommand{\ket}[1]{| #1 \rangle}
\newcommand{\bra}[1]{\langle #1 |}

\begin{document}
\title{Directional emission of single photons from small atomic samples}
\author{Yevhen~Miroshnychenko}
\affiliation{Lundbeck Foundation Theoretical
  Center for Quantum System Research\\
  Department of Physics and Astronomy\\
  University of
  Aarhus\\
  DK 8000 Aarhus C, Denmark}

\author{Uffe~V.~Poulsen}
\affiliation{Department of Engineering\\
  University of
  Aarhus\\
 DK 8200 Aarhus N, Denmark}

\author{Klaus~M{\o}lmer}
\affiliation{Lundbeck Foundation Theoretical
  Center for Quantum System Research\\
  Department of Physics and Astronomy\\
  University of
  Aarhus\\
  DK 8000 Aarhus C, Denmark}

\date{\today}
\begin{abstract}
We provide a formalism to describe deterministic emission of single photons with tailored spatial and temporal profiles from a regular array of multi-level atoms. We assume that a single collective excitation is initially shared by all the atoms in a metastable atomic state, and that this state is coupled by a classical laser field to an optically excited state which rapidly decays to the ground atomic state. Our model accounts for the different field polarization components via re-absorption and emission of light by the Zeeman manifold of optically excited states. 
\end{abstract}

\maketitle
\section{Introduction}
\label{sec:intro}
Single photons may serve as flying qubits to communicate between registers of stationary, material qubits in quantum computing architectures \cite{Olmeschenk10}, and they may be applied in protocols for quantum cryptography. In these protocols transmission losses over long distances can be counteracted by transfer of the light state qubits to quantum repeaters for purification and entanglement distillation \cite{Duan01}. Candidates for stationary qubits that can effectively interact with single photons are optically thick ensembles of atoms \cite{Hau99,Kuzmich97, Hammerer10, Schnorrberger09,Black05}, rare-earth ions in crystals
\cite{Bonarota11,Usmani10,Afzelius10,Gisin07}, vibrational excitations in diamond crystals \cite{Lee11}, as well as systems with fewer particles using optical cavities to increase the interaction with the photon field \cite{Keller04,Kuhn10,Boozer07,Albert11,Hennessy07}. The systems mentioned can provide a deterministic coupling of the material system to a suitably tailored spatial and temporal photon wave packet. There are also a number of probabilistic protocols, where measurement processes herald the successful generation of non-classical excitations of either the photon field or the medium \cite{Neergaard06,Choi10,Jiang04}.
In this article we utilize the fact that an ensemble of atoms of just few hundred atoms may interact strongly with a single mode of light with a specifically chosen mode function. We identify this mode function by calculating the emitted field from the atomic ensemble, prepared in a collectively excited state. By a time reversal argument, the complex conjugate of this emitted field may be injected on a ground state atomic ensemble and will then be fully absorbed at a definite instant of time \cite{Pedersen09,Stobinska09}. We discuss the possibility to shape the temporal profile of the emitted photon, and, in particular, the creation of time-symmetric photon wave packets, as such packets can then be emitted by one ensemble and absorbed by another one in a fully deterministic manner.
\begin{figure}[tbp]
  \centering
{\includegraphics{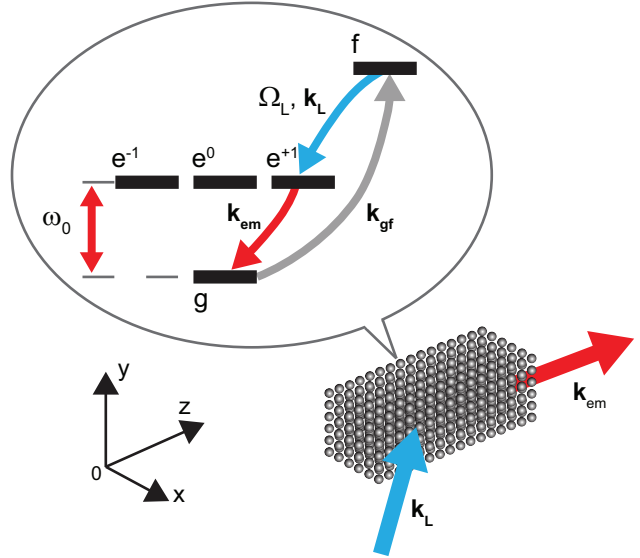}}
  \caption{Atoms with three excited substates $e^{-1}$, $e^{0}$ and $e^{+1}$, a ground state $g$ and a long lived state $f$ are arranged in a regular lattice. The state $f$ is coupled to the state $e^{+1}$ by a classical laser field. The excited states decay to the ground state emitting a photon with $\sigma^-$, $\pi$ and $\sigma^+$ polarized photons, respectively. The direction of the emission is determined by the wave number of the atomic $g-f$ coherence and the wave number of the out-coupling field, $\vec{k}_{em}=\vec{k}_{gf}-\vec{k}_L$. \label{fig:level_scheme}}
\end{figure}
We focus on samples of few hundred atoms, distributed over a few micrometer spatial extent. In such ensembles, the Rydberg blockade interaction may be used to establish singly excited states and, subsequently, single photon states \cite{Lukin01,Saffman10,Saffman02,Dudin12}, while photonic qubits, collectively absorbed by the atoms may be manipulated by Rydberg state mediated quantum gate operations \cite{Pedersen09,Brion07,Brion08}.

The collective interaction of light with ensembles of absorbers and scatterers has been an active field of study since the early days of electromagnetism, while collective phenomena in spontaneous emission received wide attention with the pioneering work on Dicke superradiance from population inverted samples \cite{Dicke53}. Early studies of collective emission from ensembles with few excitations \cite{Lehmberg70,Milloni74,Buley64,Rehler71, Manassah83} (see also \cite{Svidzhinsky08} and references therein) have been followed by a recent flourishing of analyses \cite{Svidzinsky10,Mazets07,Scully06,Pedersen09, Porras08, Bienaime12, Bariani12a,Bariani12b,Li12}, which apply a Born-Markov approximation and eliminate the field degrees of freedom to obtain coupled equations for the atomic excited state amplitudes. Approximate solutions to these equations may be derived, e.g.,  with the assumption of a scalar description of the field, but for only few hundred atoms, they may also be solved directly on a computer.

In this manuscript we generalize the previous analyses to account for the full vector character of the quantized radiation field. We establish coupled equations for excited state amplitudes on a suitable set of atomic Zeeman sub-levels, emitting and reabsorbing the different polarization components of the field, and we solve the equations numerically to identify the full temporal, spatial and polarization content of the emitted light.

In Sec. II we derive the coupled atomic equations under the Born-Markov approximation. In Sec. III, we present numerical results for the photon modes emitted by samples of atoms with different spatial geometries. In Sec. IV, we describe the use of a coupling laser field to control the temporal shape of the emitted photon wave packet, and in Sec. V, we present a brief conclusion and outlook of the work.

\section{Dipole-dipole interaction}
\label{sec:dipole}

We want to describe the experimental situation where a collection of $N$ atoms can be prepared in a single ground state $g$, and where a suitable, symmetric excitation mechanism allows the preparation of a state
\begin{equation}
|\psi\rangle = \sum_{j=1}^N a_j |g_1 g_2 ..., f_j, ... g_N\rangle
\end{equation}
with a single atom transferred to the metastable state $f$, see Fig. 1. We assume that $N$ atoms are located at the positions $\vec{r}_j$ ($j=1,..,N$). With plane wave  excitation laser fields, the amplitudes $a_j$ may have have equal magnitude and they depend on the phase of the fields at the atomic locations,  $a_j=\frac{1}{\sqrt{N}}\exp(-i\vec{k}_{gf}\cdot \vec{r}_j)$. The Rydberg blockade mechanism may restrict the system to precisely one excitation, and the state $f$, sketched in the figure, may indeed represent a long-lived Rydberg state, or a Raman process via a Rydberg state, may transfer precisely one atom to a long-lived low lying atomic state.

To release a photon from the system, we use a classical laser field with the Rabi frequency $\Omega_L$ to drive the atomic $f$-state amplitude into an optically excited state $e$, with a strong dipole coupling to the ground state $g$. The system now acts as an antenna array for dipole radiation on the transition $e-g$, and this is the cause of the desired directionality of the emitted light. As indicated in Fig.~\ref{fig:level_scheme}, the initially populated states may be extremal Zeeman sub-levels with well defined polarization selection rules, and the photon emitted on the $e-g$ transition may be $\sigma^+$ polarized with respect to the atomic quantization axis. This field, however, may be reabsorbed by another atom located in an arbitrary direction from the emitter, and here, the expansion of the field on polarization components permits excitation with selection rules $\Delta m= 0,\pm 1$. To describe the many-atom emission, we thus have to include other Zeeman sublevels than the ones initially populated. This motivates the model depicted in Fig. 1, with unique states $g$ and $f$, and three excited states $e^0,\ e^{\pm 1}$, corresponding to a $J=0\ -\ J=1$ optical transition. This configuration is the simplest extension of a two-level model atoms that allows us to fully take into account the polarization of the emitted and re-absorbed light as well as the resulting dipole-dipole interactions between the atoms.

In the following we will use the short hand notation for singly excited states of the atomic ensemble, $|f_j\rangle \equiv |g_1 g_2 ... f_j ... g_N\rangle$, and similarly for $\ket{e^{\nu}}$, with $\nu=0,\pm 1$.

In the dipole approximation the interaction of atoms with photons is described by a Hamiltonian \cite{Milloni74}:
 \begin{equation}
  \label{eq:Hamiltonian}
  \hat{H} = \hat{H}_0+\hat{H}_{int},
\end{equation}
where
\begin{multline}
\label{eq:H0}
  \hat{H}_0=  \sum_{\vec{k}} \sum_{\lambda}\hbar \omega_k a_{\vec{k} \lambda}^+ a_{\vec{k} \lambda} +\sum_{j=1}^N \sum_{\nu=-1}^1 \hbar \omega_0 \ket{e_j^\nu} \bra{e_j^\nu}+\\
+\sum_{j=0}^N\hbar \omega_{fg} \ket{f_j}\bra{f_j}
\end{multline}
is the atom-field Hamiltonian and the interaction part is
\begin{equation}
\label{eq:Hint}
  \hat{H}_{int}= \hat{H}_{L}+ \hat{H}_{V}.
\end{equation}
The semi-classical coupling to the initial long-lived state is
\begin{multline}
\label{eq:HL}
  \hat{H}_{L}= \sum_{j=1}^N \hbar \frac{\Omega_L}{2}\left[ \left(\vec{\sigma}_{fe}^j\cdot \vec{\epsilon}_L\right) e^{-i \omega_L t}+
\left(\vec{\sigma}_{ef}^j\cdot \vec{\epsilon}_L\right) e^{i \omega_L t}
 \right],
\end{multline}
where $\vec{\epsilon}_L$ is the polarization direction of the coupling field with the optical frequency $\omega_L$ and $\vec{\sigma}_{fe}^j= \hat{d}_{fe} \ket{f_j} \bra{e_j^{+1}}$. The direction of the dipole moment for this transition $\hat{d}_{fe}$ we further assume to be parallel to $\vec{\epsilon}_L$, so that the transfer of amplitude happens exclusively to the state $|e^{+1}\rangle$.

The coupling of the atomic dipole between $\ket{g}$ and $\ket{e}$ to the quantized radiation field modes is described by
\begin{multline}
\label{eq:Hv}
\hat{H}_{V}= -i \sum_{j=1}^N \sum_{\vec{k}} \sum_{\lambda}
\hbar g_k  [
\left ( \vec{\sigma}_{eg}^j\cdot \vec{\epsilon}_{\vec{k}\lambda}\right ) a_{\vec{k}\lambda} e^{i \vec{k}\cdot \vec{r}_j}-\\
-\left ( \vec{\sigma}_{ge}^j\cdot \vec{\epsilon}_{\vec{k}\lambda}\right ) a_{\vec{k}\lambda}^+ e^{-i \vec{k}\cdot \vec{r}_j}
].
\end{multline}
Here, the atomic dipole operator is defined as $\vec{\sigma}_{ge}^j=\sum_{\nu=-1}^1 \hat{d}_{g \nu}^j \ket{g} \bra{e_j^{\nu}}$, where $\hat{d}_{g \nu}^j$ is the unit vector in the direction of the corresponding dipole moment for the $\ket{g}-\ket{e^{\nu}}$ transition. $a_{\vec{k}\lambda}$ is the annihilation operator of a vacuum electro-magnetic field mode $\vec{k}$ with the polarization $\lambda$ in the direction $\vec{\epsilon}_{\vec{k}\lambda}$. The atom-field coupling strength is $g_k=d_{eg}\left(\frac{2 \pi \omega_k}{\hbar V} \right)^{1/2}$ with the quantization volume $V$ and $\omega_k=c k$.

We henceforth ignore spontaneous emission on the $e$-$f$ transition; this may on the one hand be chosen as a transition with a weaker dipole moment, and on the other hand it does not experience the collective enhancement, that we shall observe on the $e$-$g$ transition.

We expand the time dependent solution of the Schr\"{o}dinger equation for $N$ atoms and the field as a superposition of Fock states with a single atomic or photonic excitation
\begin{multline}
\label{eq:wavefunction}
\ket{\psi(t)}=\sum_{j=1}^N a_j(t) e^{-i \omega_{fg} t}\ket{f_j} \ket{g}\ket{0}+\\
+\sum_{j=1}^N \sum_{\nu=-1}^{+1}\beta_j^{\nu}(t) e^{-i \omega_0 t}\ket{0}\ket{e_j^{\nu}}\ket{0}+\\
+\sum_{\vec{k}}\sum_{\lambda}e_{\vec{k}\lambda}(t) e^{-i \omega_k t}\ket{0}\ket{g}\ket{1_{\vec{k},\lambda}},
\end{multline}
where $\ket{g}$ represents the state with all atoms in the ground state.

Note that we use the rotating wave approximation (RWA) in Eq.~(\ref{eq:HL}) because we treat this transition semi-classically. In the quantized atom-light interaction, Eq.~(\ref{eq:Hv}), the RWA shows a rather intricate interplay with the role of virtual photon processes and a seemingly coincidental equivalence of terms as discussed for the case of two-level atoms in  \cite{Milloni74,Svidzinsky10}. This holds for the terms describing atom-atom interactions. A generalization of this non-trivial discussion to the case of multilevel atoms interacting with a quantized vector field will be presented elsewhere \cite{Miroshnychenko_12_RWA}. This equivalence of terms allows us to use the RWA in our treatment of quantized atom-light interaction in Eq.~(\ref{eq:Hv}).
\begin{figure}[tbp]
  \centering
{\includegraphics{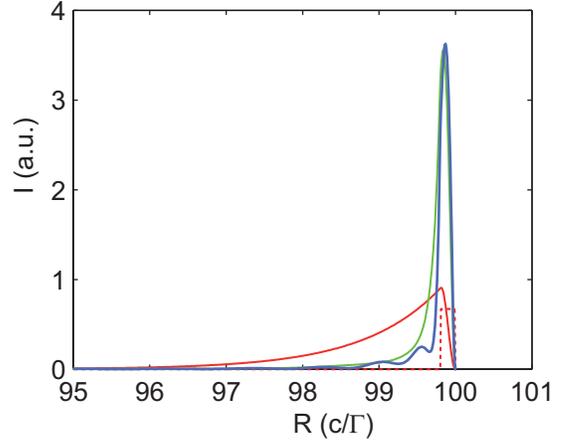}}
  \caption{A snapshot of the total intensity emitted by a cloud of $14 \times 14 \times 10$ (blue) and $3 \times 3 \times 10$ (green) atoms with the lattice spacing $d=0.25~\lambda_0$ and a single atom using a square out-coupling pulse (dotted line). Due to the coupling between different decay channels in the system, we observe beat-like behavior of the emitted light. See as well Sec.~\ref{sec:spatial} for more details. Parameters used for this simulation are $\Omega_L=8.2~\Gamma$, the single atom detuning of $\Omega_L$ light is $0$ and the pulse length $t_w=0.2~\Gamma^{-1}$.
\label{fig:beats}}
\end{figure}

Substitution of Eq.~(\ref{eq:wavefunction}) with the initial condition $e_{\vec{k}\lambda}(0)=0$ into the Schr\"{o}dinger equation with the Hamiltonian from Eq.~(\ref{eq:Hamiltonian}) yields the formal solution
\begin{multline}
\label{eq:emfield}
e_{\vec{q}\sigma}(t)=\sum_{j=1}^N \sum_{\nu=-1}^{+1} g_q e^{-i \vec{q}\cdot \vec{r_j}} \left(\hat{d}_{g \nu}^j \cdot \vec{\epsilon}_{\vec{q} \sigma }\right) \times\\
\times \int_0^t d\tau \beta_j^{\nu} (\tau) e^{-i(\omega_0-\omega_q) \tau}.
\end{multline}
Using the Markovian approximation \cite{Lehmberg70} the atomic coefficient $\beta_j^{\nu}(\tau)$ can be approximated by $\beta_j^{\nu}(t)$ in Eq.~(\ref{eq:emfield}) and taken outside the integral. This allows us to substitute the photon amplitudes by expressions involving only atomic state amplitudes, which thus obey a closed set of equations:
\begin{equation}
\label{eq:a}
\dot{a}_l=\frac{\Omega_L}{2 i}e^{i (\omega_{fe}-\omega_L)t} \beta_l^{+1},
\end{equation}
\begin{multline}
\label{eq:beta}
\dot{\beta}_l^{\eta}= \frac{\Omega_L}{2 i} e^{-i(\omega_{fe}-\omega_L)t} \delta_{\eta,1} a_l - \left( \frac{\Gamma}{2}-i \Delta_{Lamb}\right)\beta_l^{\eta}-\\
-\frac{\Gamma}{2}\sum_{j=1}^N \sum_{\nu=-1}^1 \left( 1-\delta_{l,j} \right)\left(\hat{d}_{\eta g}^l \cdot \overleftrightarrow{F}_{l,g} \cdot\hat{d}_{g \nu}^j \right) \beta_j^{\nu}.
\end{multline}
Here $\Gamma$ is the single atom decay rate from $\ket{e}$ to $\ket{g}$, and $\Delta_{Lamb}$ is the single atom Lamb shift. This term contains an infinite integral, where a suitable cut-off should be applied to yield a finite physical value \cite{Trippenbach92}. After this procedure this shift can in principle be absorbed into the definition (the measured value) of the energy of the atomic state $\ket{e}$ in Eq.~(\ref{eq:wavefunction}).

The second rank tensor $\overleftrightarrow{F}_{l,j}=\overleftrightarrow{f}(k_0 R_{l,j})- i\overleftrightarrow{g}(k_0 R_{l,j})$ with
\begin{multline}
\label{eq:Ttensor}
\overleftrightarrow{f}(k R)=\frac{3}{2}
\left( \overleftrightarrow{I}-\hat{R}\hat{R} \right )\frac{\sin(k R)}{k R}+\\
+\frac{3}{2}   \left( \overleftrightarrow{I}- 3 \hat{R}\hat{R} \right)\left( \frac{\cos(k R)}{(k R)^2}-\frac{\sin(k R)}{(k R)^3} \right),
\end{multline}
\begin{multline}
\label{eq:Ttensor2}
\overleftrightarrow{g}(k R)=\frac{3}{2}\left( \overleftrightarrow{I}-\hat{R}\hat{R} \right )\frac{\cos(k R)}{k R}-\\
-\frac{3}{2}   \left( \overleftrightarrow{I}- 3 \hat{R}\hat{R} \right)\left( \frac{\sin(k R)}{(k R)^2}+\frac{\cos( k R)}{(k R)^3} \right)
\end{multline}
and $\vec{R}_{l,j}=\vec{r_l}-\vec{r_j}$ accounts for the field mediated interaction between the atoms $l$ and $j$. Here $\overleftrightarrow{I}$ is the unity tensor and $\hat{R}\hat{R}$ is the projection onto the direction given by $\vec{R}$  \cite{Lehmberg70}.

Eqs.~(\ref{eq:a})-(\ref{eq:beta}) account for the zero-photon subspace component of the total wavefunction of the atoms and the quantized field,
\begin{equation}
\label{eq:wfuncAtomsOnly}
\ket{\Psi_0(t)}=\sum_{j=1}^N a_j(t)\ket{f_j}
+\sum_{j=1}^N \sum_{\nu=-1}^{+1}\beta_j^{\nu}(t)\ket{e_j^{\nu}},
\end{equation}
and this state component is described by an effective non-Hermitian Hamiltonian
\begin{equation}
\label{eq:Heff}
\hat{H}^{eff}=\hat{H}_0^{eff}+\hat{H}_L^{eff}+\hat{H}_{dd}^{eff}
\end{equation}
with
\begin{equation}
\label{eq:Heff_0}
\hat{H}_0^{eff}=-i \hbar \left( \frac{\Gamma}{2}-i \Delta_{Lamb} \right)\sum_{l=1}^N\sum_{\eta=-1}^{+1} \ket{e_l^{\eta}}\bra{e_l^{\eta}}
\end{equation}
describing individual single atom effects,
\begin{equation}
\label{eq:Heff_L}
\hat{H}_L^{eff}=-\frac{\hbar \Omega_L}{2} \sum_{l=1}^N e^{i (\omega_{fe}-\omega_L)t}\ket{f_l}\bra{e_l^{+1}} + H.C.
\end{equation}
giving the coupling from the long-lived state $f$ and
\begin{multline}
\label{eq:Heff_dd}
\hat{H}_{dd}^{eff}=i \hbar \frac{\Gamma}{2}\sum_{
\substack{
   j,l=1
  } }^N \sum_{
\substack{
   \nu,
   \eta=-1
  } }^{+1} \left( 1-\delta_{l,j}\right)\times \\
\times \left( \hat{d}_{\eta g}^l \cdot \overleftrightarrow{F}_{l,j}\cdot \hat{d}_{g \nu}^j \right) \ket{e_j^{\eta}} \bra{e_j^{\nu}}
\end{multline}
expressing collective dispersive and dissipative effects between the atoms in the ensemble.
\begin{figure*}[tbp]
  \centering
  \includegraphics{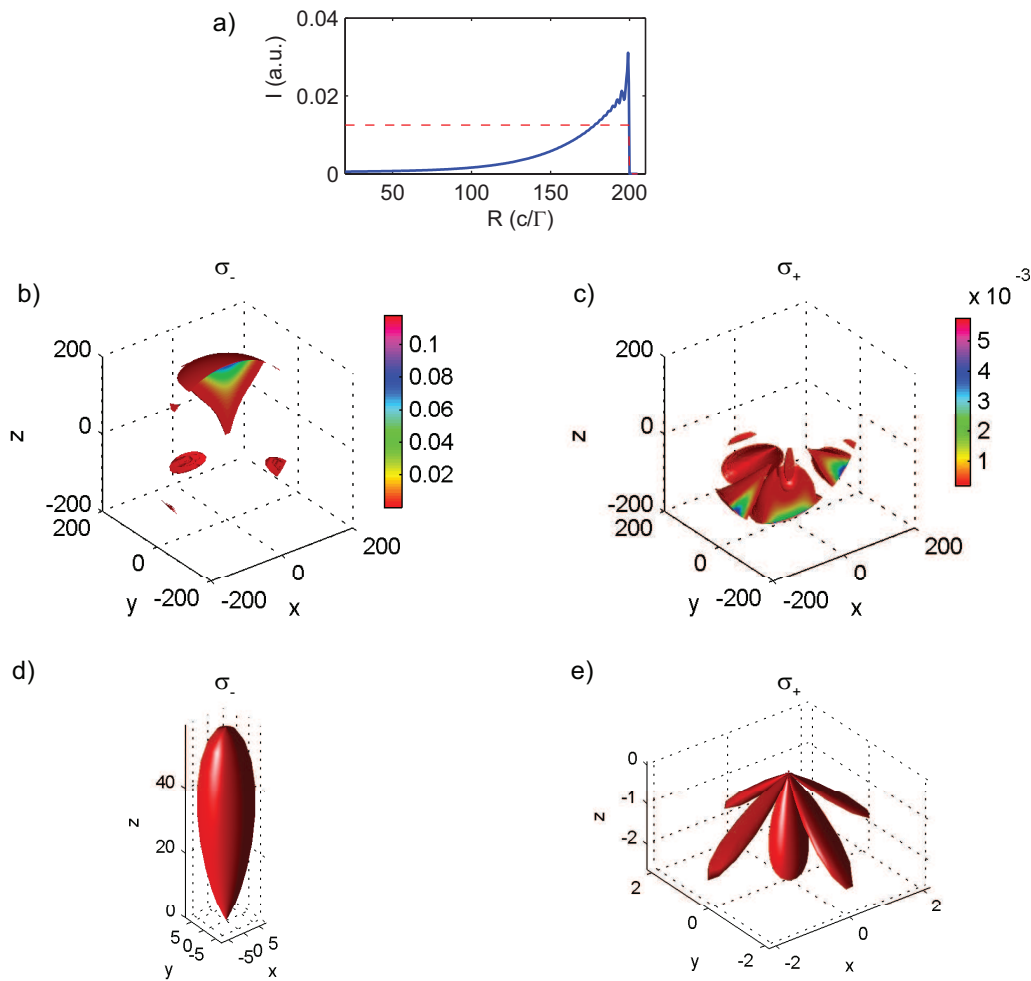}
  \caption{Emission of a single photon from a small atomic sample of $3 \times 3 \times 8$ atoms arranged in a lattice with the spacing $d=0.60~\lambda_0$. 
a) Integrated intensity as function of the distance from the atomic sample (solid line) and the out-coupling pulse shape (dotted line).
b) Spatial distribution of intensity of left-handed polarized light for a given time $t=200 ~\lambda_0/c$. The intensity scale is given in arbitrary units. The spatial coordinates are given in units of $\lambda_0$. c) Same as b), but for the opposite handedness of the light. d-e) The angular intensity distributions of light at $R=199~c/\Gamma$ corresponding to Figs. b) and c) at the moment of the maximum intensity. Note that the intensity scales for the two polarizations differ by an order of magnitude. Parameters used for this simulation are $\Omega_L=2~\Gamma$ and the single atom detuning of the $\Omega_L$ field is $10~\Gamma$.
}
  \label{fig:3d_largeSeparations}
\end{figure*}

In the rest of this paper we are interested in the spatial and temporal emission profiles. The $\overleftrightarrow{g}$-part of $\overleftrightarrow{F}$ defines the Hermitian part of $\hat{H}_{dd}^{eff}$, i.e. the coherent exchange of excitation between the atoms. The $\overleftrightarrow{f}$ part conversely defines the anti-Hermitian part corresponding to the decay by emission of light, and thus population of the one-photon quantum state component. If the Hermitian part is diagonalized, delocalized orthogonal eigenmodes with collectively "Lamb  shifted" energies are obtained. If the anti-Hermitian part is diagonalized, delocalized orthogonal independently decaying modes are obtained. Some of these modes have a decay time longer than $\Gamma$, i.e., they are  Dicke subradiant modes \cite{Dicke53}, and some decay faster than $\Gamma$, i.e., they are Dicke superradiant modes, see \cite{Pedersen09}, \cite{Svidzinsky10} and references therein. The Hermitian and the anti-Hermitian parts do not commute, and hence the eigenmodes of the full Hamiltonian $\hat{H}^{eff}$, which provide the time dependent atomic state as a single sum of complex exponentially weighted vectors, are \emph{not} orthogonal. This will result in coupling between different decay channels and a quantum beat-like behavior for the decay modes \cite{Bienaime12}. The blue line in Fig.~\ref{fig:beats} shows a snapshot of the total intensity emitted by a cloud of $14\times 14\times 10$ atoms excited from the levels $\ket{f_j}$ to the levels $\ket{e_j^{+1}}$ by a square $\Omega_L$ pulse (dotted line). The initial emission rate is faster than the single atom emission rate $\Gamma$ presented by the red line. Additionally, we see a modulation of the emitted intensity due to the coupling of the different non-orthogonal decay modes. With the reduction of the cloud size, the coupling between the decay modes becomes weaker, but the superradiance behavior remains pronounced as shown by the green line in Fig.~\ref{fig:beats} for the array of only $3\times 3\times 10$ atoms.

\section{Spatial photon modes}
\label{sec:spatial}
The eigenmode expansion of Eqs.~(\ref{eq:a})-(\ref{eq:beta}) for the atomic excitation amplitudes formally yields a solution for each individual atomic excited state amplitude as a sum of exponential functions of time with complex arguments.
Once the atomic evolution is determined, the light emission is given by the integrals in  Eq.~(\ref{eq:emfield}). We note that for the relevant time scales, longer than the ensemble excited state lifetime, these integrals involve only decaying exponential functions, and they may in practice be extended to infinity. This allows to determine the (far) field eigenmode expansion coefficients as algebraic expressions involving the mode expansion coefficients divided by the sum of the complex eigenvalues and the frequency difference appearing explicitly in Eq.~(\ref{eq:emfield}). This means that we can readily determine the field amplitudes on any chosen set of field modes after diagonalization of the atomic problem, at a cost that depends only on the number of atoms, and calculations with even thousands of atoms are realistic.

The probability to detect a photon at a position $\vec{r}$ at time instance $t$, much later than $L/c$, where $L$ is the linear sample length, is given by \cite{Scully97}
\begin{equation}
\label{eq:G1}
I_{\epsilon}\left(\vec{r},t\right)=\bra{\psi(t)} E_{\epsilon}^{(-)}(\vec{r})E_{\epsilon}^{(+)}(\vec{r})\ket{\psi(t)}.
\end{equation}
Here $\epsilon$ is the handedness of the photon, i.e., its circular polarization along the line connecting the atomic ensemble and the detector at position $\vec{r}$.
The positive frequency component of the desired polarization is given by $E_{\epsilon}^{(+)}(\vec{r})= \left( \vec{\epsilon} \cdot \vec{E}^{(+)}(\vec{r}) \right)$
with \cite{Milloni74}
\begin{equation}
\label{eq:E}
\vec{E}^{(+)}\left( \vec{r} \right)=i \sum_{\vec{q}}\sum_{\sigma} \hbar g_{q} \vec{\epsilon}_{\vec{q}, \sigma}   a_{\vec{q}, \sigma} e^{i \vec{q} \cdot \vec{r}}.
\end{equation}
See Appendix~\ref{app:Emission} for the full derivation of $I_{\epsilon}$.

In this section we assume that the atomic system is initially prepared in a so-called timed Dicke state 
\begin{equation}
\label{eq:initial_state}
a_j(0)=\frac{1}{\sqrt{N}} e^{-i \vec{r}_j \cdot \vec{k}_{gf}},
\beta_j^{\nu}(0)=0.
\end{equation}
This state is coupled to $\ket{e^{+1}}$ by switching on the laser field described by the Rabi frequency $\Omega_L$, and we first study the case where this coupling field is kept constant. In order to avoid resonant coupling to individual modes of Eq.~(\ref{eq:Heff_dd}), we assume as well that the $\Omega_L$ field detuning from the single atom resonance $\ket{f}-\ket{e^{+1}}$ is larger than the Rabi frequency $\Omega_L$, see Sec.~\ref{sec:temporal} for further discussion.

We first consider a sample of $3 \times 3 \times 8$ 
atoms arranged in a lattice  along the $x$-, $y$- and $z$- directions, respectively, with the lattice period $d \ge \lambda_0/2$. This case would correspond to trapping of alkali atoms on the $\omega_0$ transition in a red detuned lattice.

The calculated emission of a single photon from this atomic sample is presented in Fig.~\ref{fig:3d_largeSeparations}. Figure~\ref{fig:3d_largeSeparations}a shows the intensity of the field, integrated over directions, at different distances from the atomic sample. The field propagates at the speed of light, and this snapshot of the intensity distribution with distance reflects how the state $|f\rangle$ population has gradually decayed via the optically excited states since the coupling field was switched on. The most prominent feature is the overall exponentially decaying shape, but the figure also show the residual beat-like behavior. This beat-like behavior becomes even more pronounced with the increase of the array size as was shown in Fig.~\ref{fig:beats} and is in agreement with our discussion of the real and imaginary eigenvalues of the problem. Figures~\ref{fig:3d_largeSeparations}b and c show the spatial intensity distribution, indicated by colors, for the same parameters as in part a of the figure. Figure~\ref{fig:3d_largeSeparations}b shows the intensity distribution for the polarization of the emitted light, which is expected to be dominant for the level scheme by the dipole selection rule. We observe that most of the light is emitted in a narrow forward peak. Figure~\ref{fig:3d_largeSeparations}c shows that in the backward direction, a small component (note the different scale) is emitted with the opposite polarization, again in agreement with the atomic dipole selection rule. To quantify the angular distribution of the emitted light, Figs.~\ref{fig:3d_largeSeparations}d,e show polar plots of the intensity of the left-handed and right-handed light components at the distance $R=199~c/\Gamma$.

\begin{figure*}[tbp]
  \centering
{
\includegraphics{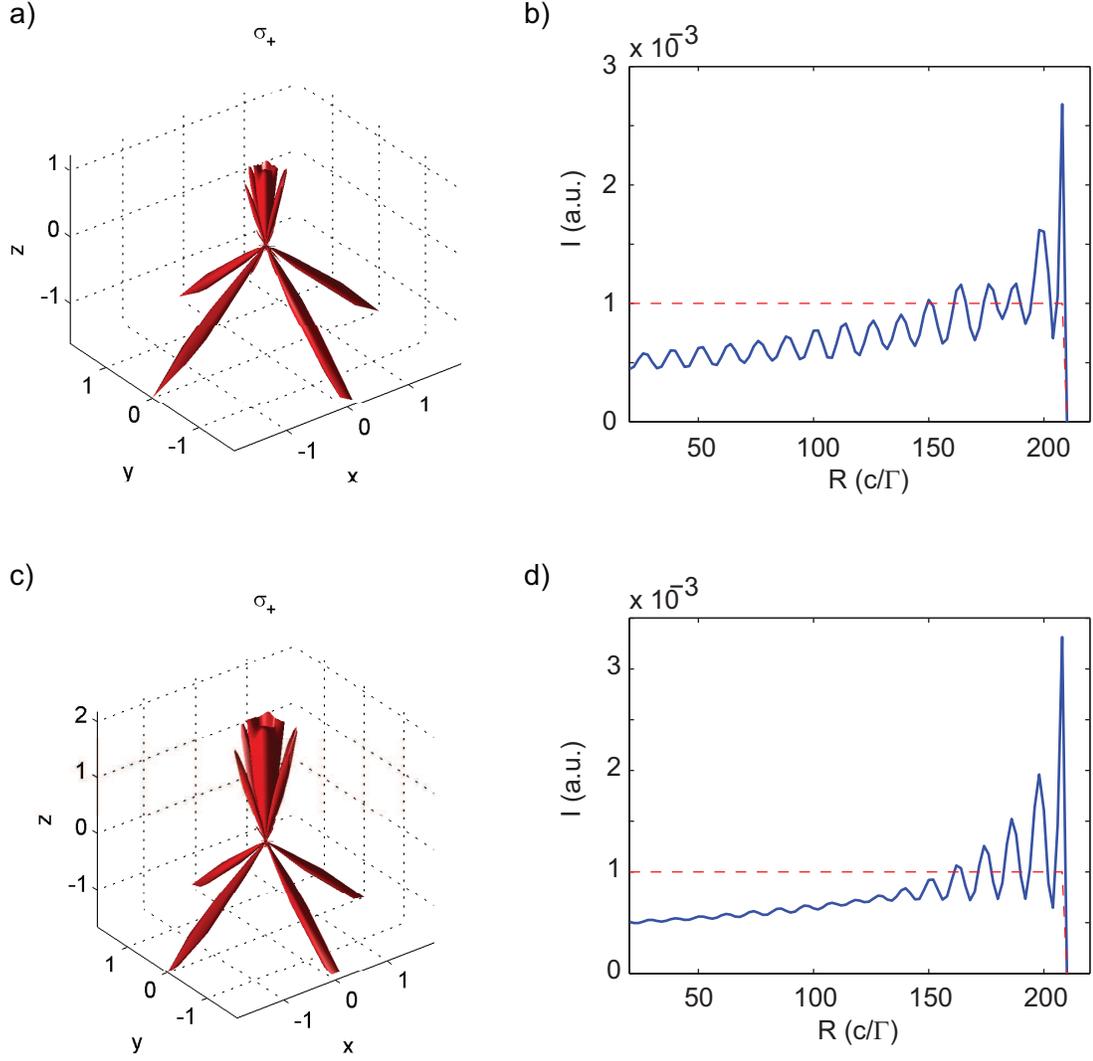}}
  \caption{Frustrated emission from a sample of $8 \times 8 \times 8$ with the lattice spacing $d=0.60~\lambda_0$. a) A snapshot of the angular distribution at $R=208~c/\Gamma$ and b) angularly integrated intensity of the emitted light for the case of all $\ket{e^{-1}}$, $\ket{e^{0}}$ and $\ket{e^{+1}}$ levels are included. c)-d) The corresponding graphs, if only $\ket{e^0}$ level is included. The coupling laser parameters are the same as in Fig.~\ref{fig:3d_largeSeparations}.
\label{fig:frustrated}}
\end{figure*}

The directionality of the emitted light as a function of the lattice spacing  was analytically studied by Porras and Cirac \cite{Porras08} for the case of two-level atoms. In this model the radiative pattern can be factorized into two parts: 1) collective scalar field radiative effects and 2) a dipole radiative pattern of individual atoms. In a general case of multiple exited states and a vector electro-magnetic field, atomic reabsorption of photons with a resulting redistribution of angular momentum associated with the directions of propagation between the atoms, does not justify such a simple factorization of the radiation pattern and a scalar field approximation. Our calculations show, that these reabsorption effects become important for the case of strongly coupled atoms, i.e., for atomic spacings $d \apprle 0.25~\lambda_0$, and in the case of frustrated forward emission even for spacings $d \approx \lambda_0$, see Fig.~\ref{fig:frustrated}. In both cases the subradiant modes play an important role.

Although the detailed analysis of these cases is beyond the scope of the current work, we give short example for the case of frustrated emission. We consider emission by an array of $8 \times 8 \times 8$ atoms with the spacing $d=0.60~\lambda_0$. The atoms are initially prepared in the superposition timed Dicke state of the $\ket{f}$ levels Eq.~(\ref{eq:initial_state}) as in all previous examples, favoring the emission in the $z$-direction, but the storage level $\ket{f}$ is coupled now to the short lived state $\ket{e^0}$. This state can directly decay to the ground state with the emission of a $\pi$-polarized photon, relative to the $z$-axis, see Fig.~\ref{fig:level_scheme}. Since the angular emission pattern of this transition has zero intensity in the $z$-direction, which is the preferred emission direction of this atomic sample, we arrive at the conflict between the preferred and allowed emissions. An example of the frustrated emission is presented in Fig.~\ref{fig:frustrated}. Here we plot a snapshot of the angular distribution at $R=208~c/\Gamma$ and the integrated intensity emitted by an array of $8 \times 8 \times 8$ atoms with $d=0.60~\lambda_0$ and a long coupling pulse $\Omega_L$ (dotted line). Figures~\ref{fig:frustrated}a-b show the case, where the coupling of $\ket{g}$ to all $\ket{e^{-1}}$, $\ket{e^{0}}$ and $\ket{e^{+1}}$  states is allowed. Fig.~\ref{fig:frustrated}c-d show the situation of emission from two level atoms, where only $\ket{g}$ and $\ket{e^{0}}$ are coupled. Both the angular dependence and the time evolution show differences in these two situations. This difference stems from the reabsorption of virtual photons with different polarizations.

\begin{figure*}
  \centering
  \includegraphics{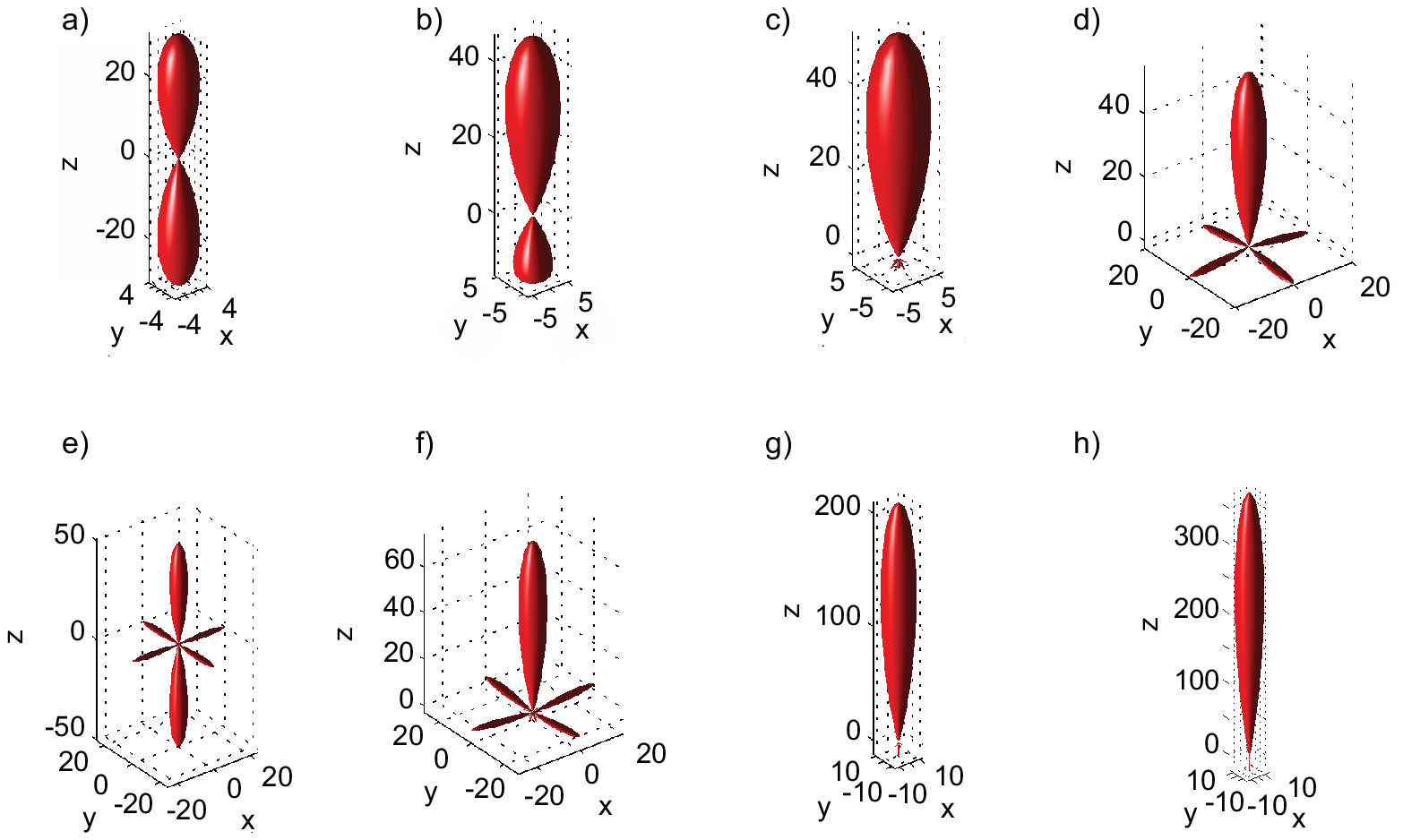}
  \caption{Total angular distribution of emitted light for a lattice $3 \times 3 \times 8$ atoms for the cases a) $d=0.50~\lambda_0$, b) $d=0.53~\lambda_0$, c) $d=0.60~\lambda_0$, d) $d=0.90~\lambda_0$, e) $d=1.00~\lambda_0$, f) $d=1.10~\lambda_0$. The total angular distribution for lattices g) $6 \times 6 \times 8$ and h) $8 \times 8 \times 8$ with $d=0.60~\lambda_0$.
The coupling laser parameters are the same as in Fig.~\ref{fig:3d_largeSeparations}.
}
  \label{fig:angular}
\end{figure*}
Our calculations also show, that the analysis of \cite{Porras08} is qualitatively correct for atoms with larger separation.
The directionality of emission in this case depends on the lattice spacing and can be understood as interference of Bragg scattering contributions. For a critical spacing of $d=\lambda_0/2$ or an integer multiple of $\lambda_0/2$, the photon is emitted in the forward and backward directions with equal probability, see Fig.~\ref{fig:angular}a and e. With an increased value of $d$ this symmetry is broken and the forward emission peak becomes dominant, see Fig.~\ref{fig:angular}b-d and f. Due to the diffraction-like effects the directionality of the emitted light improves with the increase of the array size. Figures~\ref{fig:angular}c, g and h show the angular distribution of emitted light for lattices with the same spacing, but with an increasing number of atoms.\\

In this section we have studied the collective emission of single photons and we have analyzed details of their directional distribution and their associated polarization properties. In the next section we turn to the temporal shape of the emitted light pulses and to the experimental means to control the light mode of the emitted photon.

\section{Temporal photon modes}
\label{sec:temporal}

The angular emission patterns shown in the previous section were all
calculated assuming a constant intensity of the laser which drives the $f
\leftrightarrow e$ transition. This leads to a highly asymmetric temporal profile of the emitted light shown in Fig.~\ref{fig:3d_largeSeparations}a.
By controlling the temporal profile of the coupling $\Omega_L(t)$ one can control the temporal shape of the emitted light, and this can be used to transfer the atomic excitation from the state $\ket{f}$ to, e.g., a temporary symmetric emitted photon wave packet. Such a wave packet can be reabsorbed in a second atomic ensemble, if one employs the time reversed control field. Therefore we may imagine a collection of atomic ensembles as quantum repeater stations, where photon pulses are absorbed and reemitted, possibly after suitable entanglement distillation and state purification \cite{Gisin07,Duan01}.

To design appropriate outcoupling laser fields, we start with a system prepared in the states $\ket{f_j}$ described by Eq.~(\ref{eq:initial_state}) with a vanishing coupling $\Omega_L$. By gradually increasing the coupling strength we transfer population to the states $\ket{e_j^{+1}}$, which decay to the ground state by emission of light. The intensity of the emitted light is given by the population of the excited state $\ket{e_j^{+1}}$, and our goal is thus to control this population.

Due to the complexity of the non-orthogonal eigenmodes of the coupling Hamiltonian, see Sec.~\ref{sec:dipole}, we focus here on the conceptually easiest strategy, which is an \textit{adiabatic} out-coupling. By introducing a detuning $\delta=\omega_{fe}-\omega_L$ and $\beta_l^{\nu}=\tilde{\beta}_l^{\nu}e^{-i \delta t}$ we can rewrite Eq.~(\ref{eq:a})-(\ref{eq:beta})

\begin{equation}
\label{eq:a_tilde}
\dot{a}_l=\frac{\Omega_L}{2 i} \tilde{\beta}_l^{+1},
\end{equation}
\begin{multline}
\label{eq:beta_tilde}
\dot{\tilde{\beta}}_l^{+1}= \frac{\Omega_L}{2 i} a_l+ i\left(\delta +\Delta_{Lamb} \right)\tilde{\beta}_l^{+1}-\frac{\Gamma}{2}\tilde{\beta}_l^{+1}-\\
-\frac{\Gamma}{2}\sum_{j=1,j\ne l}^N \left(\hat{d}_{\eta g}^l \cdot \overleftrightarrow{F}_{l,g} \cdot\hat{d}_{g \nu}^j \right) \tilde{\beta}_j^{+1}.
\end{multline}
Since the state $\ket{f_j}$ is coupled by a strong field $\Omega_L$ to  $\ket{e_j^{+1}}$, whereas the substates $\ket{e_j^{-1}}$ and $\ket{e_j^{0}}$ are only coupled by virtual photon processes to the state $\ket{e_j^{+1}}$, we have neglected the populations of the $\ket{e_j^{-1}}$ and $\ket{e_j^{0}}$ substates. Assuming further that the detuning $\delta$ is much larger than the Rabi frequency $\Omega_L$, the single atom decay rate $\Gamma$ and the Lamb shifts, we adiabatically eliminate the state $\tilde{\beta}_l^{+1}$

\begin{multline}
\label{eq:beta_adiabatic}
\tilde{\beta}_l^{+1}=\left(\frac{\Omega_L}{2 \delta} - i \frac{\Gamma}{2 \delta} \frac{\Omega_L}{2 \delta}\right) a_l-\\
-i\frac{\Gamma}{2 \delta}\sum_{j=1,j\ne l}^N \left( \hat{d}_{+1,g}^l\cdot \overleftrightarrow{F}_{l,j} \cdot \hat{d}_{g,+1}^j \right) \tilde{\beta}_j^{+1}.
\end{multline}
We can see directly that the leading term of $\tilde{\beta}_l^{+1}$ is of the order $O\left(\frac{\Omega_L}{\delta}\right)$, since $a_l$ is of the order of one. Therefore the population of the state $\ket{e_l^{+1}}$ is always much smaller than unity and quickly decays to the ground state. Hence the intensity of the emitted light is approximately given by the population of the state $\ket{e_l^{+1}}$. Substituting Eq.~(\ref{eq:beta_adiabatic}) into Eq.~(\ref{eq:a_tilde}) and neglecting terms of order higher than $O\left( \frac{\Gamma^3}{\delta^2}\right)$ we arrive at a closed set of equations for the $a$ coefficients:
\begin{multline}
  \label{eq:a_adiabatic}
 \dot{a}_l= -i \Delta_{light}~a_l - \frac{\Gamma'}{2} a_l - \\
-\frac{\Gamma'}{2} \sum_{j=1,j\ne l}^N \left( \hat{d}_{+1,g}^l\cdot \overleftrightarrow{F}_{l,j} \cdot \hat{d}_{g,+1}^j \right) a_j^{+1}
\end{multline}
with the light shift $\Delta_{light}=\frac{\Omega_L^2}{4 \delta}$ and the effective
decay rate $\Gamma'=\Gamma \frac{\Omega_L^2}{4 \delta^2}$ of the metastable state $\ket{f_l}$ .

Similar to Eq.~(\ref{eq:beta_tilde}), the first line of this equation describes the dynamics of an isolated atom. The second line, which depends on the geometry, is the contribution from the interactions with all other atoms of the sample via virtual photon exchange processes. Since the collective contributions both in Eq.~(\ref{eq:beta_tilde}) and in Eq.~(\ref{eq:a_adiabatic}) have the same form, and the temporal evolution of the population $\tilde{\beta}_l^{+1}$ follows $a_l$ as
\begin{equation}
\label{eq:beta_adiabatic_leading}
\tilde{\beta}_l^{+1}=\frac{\Omega_L}{2 \delta} a_l,
\end{equation}
up to the order $O\left( \frac{\Omega_L}{\delta}\right)$, we conclude that the
spatiotemporal mode function occupied by emitted photon has the same structure, up to a radial scaling factor, as in the case studied in Sec.~\ref{sec:spatial}.

In order to tailor the temporal dynamics of the emitted light, we now allow temporal modulation of the control field $\Omega_L(t)=\Omega_{L0} f(t)$ with $0\le f (t) \le 1$ for all times. With this notation Eq.~(\ref{eq:a_adiabatic}) becomes
\begin{multline}
  \label{eq:a_adiabatic_timedependent}
 \frac{1}{f(t)^2}\frac{da_l(t)}{dt}= -i \frac{\Omega_{L0}^2}{4 \delta}~a_l(t) - \frac{\Gamma}{2}  \frac{\Omega_{L0}^2}{4 \delta^2}~a_l(t) - \\
-\frac{\Gamma}{2}  \frac{\Omega_{L0}^2}{4 \delta^2} \sum_{j=1,j\ne l}^N \left( \hat{d}_{+1,g}^l\cdot \overleftrightarrow{F}_{l,j} \cdot \hat{d}_{g,+1}^j \right) a_j^{+1}(t).
\end{multline}
\begin{figure*}[tbp]
  \centering
{\includegraphics{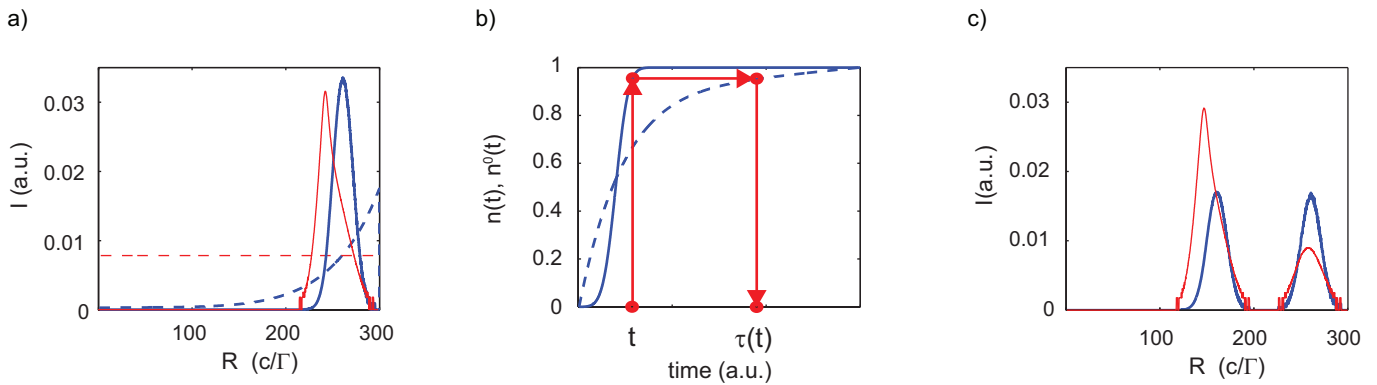}}
  \caption{Temporal shaping of the photon emission from an atomic array of $3\times 3\times 8$ atoms with $d=0.60~\lambda_0$. a) Output intensity (blue dashed line) using a constant pulse (thin, red dashed line) with $\Omega_{L0}=10.5~\Gamma$  and $\delta=120~\Gamma$. With a specially designed outcoupling pulse with peak Rabi frequency $\Omega_{L0}=42.0~\Gamma$ and the temporal shape $f(t)$ (thin red line), we obtain a Gaussian temporal shape at $R=260~c/\Gamma$ of the emitted light intensity(blue line). b) Integrated intensities $n^0(t)$ (dashed) and $n(t)$ (thick solid curve) and a graphical solution (red arrows) of the equation $n(t)=n^0(\tau(t))$.  c) Outcoupling of a double peaked pulse (thick blue line) using $\Omega_{L0}=38.85~\Gamma$ and $f(t)$ represented by the thin red line.
}
  \label{fig:adiabatic}
\end{figure*}
In contrast to a resonant outcoupling case described by Vasilev et.al. \cite{Vasilev10}, in our adiabatic outcoupling case there is no general analytic solution of this equation. Nevertheless we can find a connection between the solution $a_l^0(t)$ of Eq.~(\ref{eq:a_adiabatic_timedependent}) with a constant $\Omega_L$, i.e. $f=1$, and the solution $a_l(t)$ for an arbitrary $f(t)$. Defining
\begin{equation}
\label{eq:tau}
\tau(t)=\int_0^t f(t')^2 dt',
\end{equation}
we observe that
\begin{equation}
\label{eq:b}
b_l(\tau(t))=a_l(t),
\end{equation}
obeys the equation
\begin{multline}
  \label{eq:b_adiabatic_tau}
 \frac{db_l(\tau)}{d\tau}= -i \frac{\Omega_{L0}^2}{4 \delta}~b_l(\tau) - \frac{\Gamma}{2}  \frac{\Omega_{L0}^2}{4 \delta^2}~b_l(\tau) - \\
-\frac{\Gamma}{2}  \frac{\Omega_{L0}^2}{4 \delta^2} \sum_{j=1,j\ne l}^N \left( \hat{d}_{+1,g}^l\cdot \overleftrightarrow{F}_{l,j} \cdot \hat{d}_{g,+1}^j \right) b_j(\tau).
\end{multline}
Since this equation is identical to Eq.~(\ref{eq:a_adiabatic_timedependent}) for constant~$f$, we directly obtain its formal solution  $b_l(\tau)=a_l^0(\tau)$, and hence, the general solution to Eq.~(\ref{eq:a_adiabatic_timedependent}) reads
\begin{equation}
\label{eq:a_arbitrary}
a_l(t)=a_l^0\left(\tau(t)\right).
\end{equation}
\emph{I.e.}, the time evolution of $a_l(t)$ can be advanced or retarded with respect to $a_l^0(t)$ by choosing the appropriate function $f(t)$ in (\ref{eq:tau}).

To calculate the temporal pulse shape $f(t)$ leading to any desired output intensity $I(t)$, we observe that the number of photons emitted under constant amplitude driving is related to the population in the initial atomic state
\begin{equation}
\label{eq:n0}
n^0(t) \equiv \int_0^t dt' I^0(t')) =1-\sum_{l=1}^N|a_l^0(t)|^2,
\end{equation}
while the corresponding number of photons in the desired field is
\begin{equation}
\label{eq:n}
n(t)\equiv \int_0^t dt' I(t') = 1-\sum_{l=1}^N|a_l(t)|^2.
\end{equation}
Eq.~(\ref{eq:a_arbitrary}) implies the equation for $\tau(t)$:
\begin{equation}
\label{eq:ntau}
n(t)=n^0(\tau(t)).
\end{equation}
Fig.~\ref{fig:adiabatic}a shows the emitted intensities (dashed blue curve) obtained with a constant outcoupling amplitude (red, dashed curve) and a Gaussian pulse with $1/e$ width $15~\Gamma^{-1}$ at $R=260~c/\Gamma$ (dark blue curve) obtained with the calculated amplitude $f(t)$ (thin red curve), respectively. The corresponding integrated intensities are shown in Fig.~\ref{fig:adiabatic}b, which also illustrates the numerical procedure to solve (\ref{eq:ntau}) and establish the   correspondence between the values $\tau(t)$ and $t$, as indicated by the red arrows. By numerically differentiating the function $\tau(t)$ and using Eq.~(\ref{eq:tau}) we directly obtain the temporal profile of the outcoupling pulse $f(t)$ leading to the desired $I(t)$. This function is shown as a red curve in Fig.~\ref{fig:adiabatic}a; the spikes are due to precision errors in the calculation of the derivative of $\tau(t)$. We insert this outcoupling pulse into our simulation of the full coupled equations and we calculate the resulting emitted intensity which is, indeed, the result, presented as the blue line in Fig.~\ref{fig:adiabatic}a. To demonstrate our ability to couple out arbitrary pulse shapes, we aim in Fig.~\ref{fig:adiabatic}c at a double peaked pulse with equal heights at $R=160~c/\Gamma$ and $R=260~c/\Gamma$. The corresponding profile of $f(t)$ is shown as a red line in Fig.~\ref{fig:adiabatic}c, and again, the solid, blue curve shows the outcome of the simulation of the full set of coupled atomic equations.

We presented here a recipe for photon shaping starting from a constant amplitude control pulse. Using this as a reference, it is possible to relate the outcome of different control pulses by a suitable parametrization of the time argument, and to use the above method with experimentally measured intensities. While we assumed a varying amplitude but a constant phase of the control field $\Omega_L (t)$, to produce more complicated single photon wave packets, it may be worthwhile to study also complex valued $f(t)$.

\section{Conclusion}
\label{sec:conclusion}
We studied correlated spontaneous emission from small arrays of atoms. Our atomic model is the simplest generalization of a two-level atom, which allows to fully account for polarization effects. Numerically solving the equations of motion we demonstrated the possibility to deterministically generate single photons with a well defined spatial emission profile, which can be controlled by the geometry of the array. Finally, we presented a simple method for temporal shaping of the emitted photon, which is a prerequisite for high fidelity interfacing photon wave packets to atomic systems. Specifically we presented a recipe for generating symmetric photon wave packets, as they can be efficiently re-absorbed by another similar atomic sample.

\section*{Acknowledgments}
We acknowledge financial support from the EU Integrated Project AQUTE and from the project MALICIA under FET-Open grant number 265522.

\appendix
\section{}
\label{app:Emission}
We derive the explicit dependence of $I_{\epsilon}\left(\vec{r},t\right)$ on the atomic part $\beta$ of the wave function. We rewrite first Eq.~(\ref{eq:G1}) as \cite{Scully97}:
\begin{equation}
\label{eq:app_G1}
I_{\epsilon}=\bra{\psi} E_{\epsilon}^{(-)}\ket{0}\ket{g}\ket{0} \bra{0}\bra{g}\bra{0}E_{\epsilon}^{(+)}\ket{\psi}.
\end{equation}
The photon "wave function" is
\begin{multline}
\label{eq:app_photon_wf_initial}
\ket{\Psi_{\epsilon}}=\bra{0}\bra{g}\bra{0}E_{\epsilon}^{(+)}\left(\vec{r}\right)\ket{\psi(t)}=\\
=i \sum_{\vec{k}} \sum_{\lambda} \hbar g_k \left(\vec{\epsilon} \cdot \vec{\epsilon}_{\vec{k},\lambda}\right) e_{\vec{k},\lambda} e^{i(\vec{k} \cdot \vec{r}-\omega_k t)}
\end{multline}
with $e_{\vec{k},\lambda}$ from Eq.~(\ref{eq:emfield}).

As the second step, we simplify this expression. The summation over polarizations is performed using a recipe from the Appendix of Smith et.al. \cite{Smith91}:
\begin{multline}
\label{eq:app_photon_wf_sum_k}
\ket{\Psi_{\epsilon}}=
i \sum_{\vec{k}} \sum_{j=1}^N \sum_{\nu=-1}^{1} \left(\vec{\epsilon} \cdot (\overleftrightarrow{I}-\hat{k} \hat{k}) \cdot \hat{d}_{j,\nu}^j\right) \times\\
\times \hbar g_k^2 e^{i(\vec{k} \cdot \vec{R}_j-\omega_k t)} \int_0^t d \tau \beta_j^{\nu}(\tau) e^{-i(\omega_0-\omega_k)\tau}
\end{multline}
with $\vec{R}_j=\vec{r}-\vec{r}_j$. After replacing summation over the wave-vectors $\sum_{\vec{k}}$ by an integral $\frac{V}{(2\pi c)^3} \int_0^\infty d\omega_k \omega_k^2 \int d \Omega(k)$ we get
\begin{multline}
\label{eq:app_integral}
\ket{\Psi_{\epsilon}}=
\sum_{j=1}^N \sum_{\nu=-1}^{1}  i \frac{4 \pi \hbar  V}{(2\pi c)^3}\int_0^\infty d\omega_k \omega_k^2 g_k^2 e^{- i \omega_k t}\times \\
\times \int_0^t d \tau \beta_j^{\nu}(\tau) \times e^{-i(\omega_0-\omega_k)\tau} \times\\
\times \int d \Omega(k)
\left(\vec{\epsilon} \cdot (\overleftrightarrow{I}-\hat{k} \hat{k}) \cdot \hat{d}_{j,\nu}^j\right) e^{i\vec{k} \cdot \vec{R}_j}.
\end{multline}
The angular integration can already be performed at this step \cite{Smith91} resulting in
\begin{multline}
\label{eq:app_tau}
\ket{\Psi_{\epsilon}}=
\sum_{j=1}^N \sum_{\nu=-1}^{1}  i \frac{4 \pi \hbar  V}{(2\pi c)^3}\int_0^\infty d\omega_k \omega_k^2 g_k^2 e^{- i \omega_k t}\times \\
\times \int_0^t d \tau \beta_j^{\nu}(\tau)  e^{-i(\omega_0-\omega_k)\tau} \times\\
\times 4\pi \left(\vec{\epsilon} \cdot \overleftrightarrow{\zeta} (k R_j) \cdot \hat{d}_{j,\nu}^j\right)
\end{multline}
with $\overleftrightarrow{\zeta} (k R)= \left(\overleftrightarrow{I} -\hat{R}\hat{R}\right)\frac{sin(kR)}{kR}+O\left(\frac{1}{(k R)^2} \right)$ \cite{Smith91}. Therefore the photon intensity up to the order of $O\left((k R)^{-2} \right)$ with $r>>r_j$ is
\begin{multline}
\label{eq:app_k_integral}
\ket{\Psi_{\epsilon}}=
\sum_{j=1}^N \sum_{\nu=-1}^{1}  B_{\hat{r},j}^{\nu} \frac{1}{R_j} \int_0^t d \tau \beta_j^{\nu}(\tau)  e^{-i c k_0 \tau} \times \\
\times \int_0^\infty d k k^2 \left[ e^ { i k R_j+i c k ( \tau-t)} -e^ { -i k R_j+i c k ( \tau-t)}\right],
\end{multline}
where we have introduced $B_{\hat{r},j}^{\nu} =i \frac{8 \pi^2 c^4 d_{eg}}{(2 \pi c)^3}\left(\vec{\epsilon} \cdot (\overleftrightarrow{I}-\hat{R_j} \hat{R_j}) \cdot \hat{d}_{j,\nu}^j\right).$
Since during the emission the value of $k$ is peaked around the atomic resonance $k_0$ where the last time integral is relevant, we can replace $k^2$ with $k_0^2$ and extend the lower integral limit to $-\infty$ \cite{Scully97}, i.e. the Weisskopf-Wigner approximation. Using the definition of $\delta$-function $\delta(t)=\frac{1}{2\pi}\int_{-\infty}^{\infty}d k e^{i k t}$, Eq.~(\ref{eq:app_k_integral}) becomes
\begin{multline}
\label{eq:app_tau_integral}
\ket{\Psi_{\epsilon}}=
\sum_{j=1}^N \sum_{\nu=-1}^{1}  B_{\hat{r},j}^{\nu} \frac{2\pi}{c}\frac{k_0^2}{R_j}\times \\
\times  [ \int_0^t d \tau \beta_j^{\nu}(\tau)  e^{-i c k_0 \tau} \delta \left(t-\frac{R_j}{c}-\tau\right)-\\
- \int_0^t d \tau \beta_j^{\nu}(\tau)  e^{-i c k_0 \tau} \delta \left( t+\frac{R_j}{c}-\tau \right) ].
\end{multline}
Since the last integral with the delta function is always zero, we arrive at
\begin{multline}
\label{eq:app_photon_wf_final}
\ket{\Psi_{\epsilon}}=
\sum_{j=1}^N \sum_{\nu=-1}^{1}  \frac{d_{eg}k_0^2}{R_j}
 \beta_j^{\nu}\left(t-\frac{R_j}{c} \right)  e^{-i c k_0 \left(t-\frac{R_j}{c}\right)} \times \\
\times\left(\vec{\epsilon} \cdot (\overleftrightarrow{I}-\hat{R}_j \hat{R}_j) \cdot \hat{d}_{g,\nu}^{j}\right) .
\end{multline}

Finally, we arrive at the desired explicit dependence of the photon intensity on the atomic part $\beta$:
\begin{multline}
\label{eq:app_intensity_final}
I_{\epsilon}\left(\vec{r},t\right)=
\bra{\Psi_{\epsilon}} \Psi_{\epsilon}\rangle
=\frac{d_{eg}^2 k_0^4}{r^2} \times \\
\times\sum_{j,j'=1}^N \sum_{\nu,\sigma=-1}^{1}
\beta_j^{\nu}\left(t-\frac{r}{c}\right) \left( \beta_{j'}^{\sigma}\left(t-\frac{r}{c}\right)\right)^*
 e^{i k_0 \left( \hat{r}\cdot (\vec{r}_j-\vec{r}_{j'}) \right)} \times \\
\times  \left(\vec{\epsilon} \cdot (\overleftrightarrow{I}-\hat{R}_j \hat{R}_j) \cdot \hat{d}_{g,\nu}^{j}\right)
\left(\vec{\epsilon} \cdot (\overleftrightarrow{I}-\hat{R}_{j'} \hat{R}_{j'}) \cdot \hat{d}_{g,\sigma}^{j'~*}\right),
\end{multline}
where we have used $R_j\approx r$ for $\beta$ and in the denominator, but kept the significant term $R_j=r-\left(\hat{r} \cdot \vec{r}_j\right)$ in the exponents.

This equation allows a very transparent physical interpretation for the case of many noninteracting atoms. We suppose the states $\ket{f_j}$ are fully mapped onto the corresponding $\ket{e^{+1}_j}$, i.e. $a_j=0$ in Eq.~(\ref{eq:wavefunction}). Therefore the atomic evolution is described by
\begin{equation}
\label{eq:app_noninteracting_initial_cond}
\beta_j^{+1}(t)=\frac{1}{\sqrt{N}}e^{-i \vec{k}_{em}\cdot \vec{r}_j} e^{-\frac{\Gamma}{2}t},
\end{equation}
$$ \beta_j^{0}(t)=0,$$ and
$$\beta_j^{-1}(t)=0,$$
which is the solution of Eq.~(\ref{eq:beta}) for $k_0 R_{l,j}\gg 1$, i.e. vanishing coupling $\overleftrightarrow{F}_{l,j}$.
The corresponding emitted intensity is then
\begin{multline}
\label{eq:app_noninteracting_general}
I_{\epsilon}\left(\vec{r},t\right)
=\frac{d_{eg}^2 k_0^4}{N r^2} e^{-\Gamma(t-\frac{r}{c})}\sum_{j=1}^N  C_{j,j}^{\epsilon} +\\
+\frac{d_{eg}^2 k_0^4}{N r^2} e^{-\Gamma(t-\frac{r}{c})}\sum_{j=1}^N  \sum_{j'=1}^N (1-\delta_{j,j'})C_{j,j'}^{\epsilon} e^{i(k_0 \hat{r}-\vec{k}_{em}) \cdot (\vec{r}_j-\vec{r}_{j'})}
\end{multline}
with
\begin{multline}
\label{eq:app_dipole_pattern}
C_{j,j'}^{\epsilon}=\\
=\left(\vec{\epsilon} \cdot (\overleftrightarrow{I}-\hat{R}_j \hat{R}_j) \cdot \hat{d}_{g,+1}^{j~*}\right) \left(\vec{\epsilon} \cdot (\overleftrightarrow{I}-\hat{R}_{j'} \hat{R}_{j'}) \cdot \hat{d}_{g,+1}^{j'~*}\right).
\end{multline}
Equation~(\ref{eq:app_dipole_pattern}) can be further simplified using $\hat{R_j}\hat{R_j}=\hat{r}\hat{r}\left(1+ O\left( \frac{r_j}{r}\right) \right),$ and assuming that all dipoles are polarized in the same direction, i.e. $\hat{d}_{g,+1}^j=\hat{d}_{g,+1}$. In this case the function $C_{j,j}^{\epsilon}$ has an interpretation as the angular dependence of the dipole emission pattern for a given helicity $\epsilon$ expressed in a tensor form. In particular, if we introduce spherical coordinate system along the $\hat{d}_{g,0}$ direction and sum over all polarizations, we arrive at the well known dipole pattern angular dependence $\frac{1}{2} (1+cos^2\theta)$ for a dipole emitting on the $\ket{e^{+1}} \rightarrow \ket{g}$ transition.

Correspondingly, in the direction $\hat{r}=\frac{\vec{k}_{em}}{k_0}$ we have
\begin{equation}
\label{eq:app_noninteracting_directed}
I_{\epsilon}\left(\vec{r},t\right)
\approx N^2\times \frac{d_{eg}^2 k_0^4}{N r^2} e^{-\Gamma(t-\frac{r}{c})} C_{j,j}^{\epsilon},
\end{equation}
while for all other directions $\hat{r}\neq \frac{\vec{k}_{em}}{k_0}$ the exponents in Eq.~(\ref{eq:app_noninteracting_general}) average out and give
\begin{equation}
\label{eq:app_noninteracting_not_directed}
I_{\epsilon}\left(\vec{r},t\right)
\approx 1 \times \frac{d_{eg}^2 k_0^4}{N r^2} e^{-\Gamma(t-\frac{r}{c})} C_{j,j}^{\epsilon},
\end{equation}
i.e. $1/N^2$ reduced emission intensity. This is the well known mechanism of \textit{ directed emission} from an atomic sample. Note as well that in this example, and in the more general case of spin polarized two level atoms, i.e. no coupling of $\nu=+1$ states to the other $\nu=0,-1$ states, the single atom dipole emission pattern factorises out in the intensity formula Eq.~(\ref{eq:app_intensity_final}) as was pointed out by Porras and Cirac \cite{Porras08}. Nevertheless this is not true in the general case considered here due to coupling of the $\nu=0,-1$ levels to the level $\nu=+1$ via virtual photons \cite{Miroshnychenko_12_RWA}.


\end{document}